\documentclass[aps,prx,amsmath,amssymb,floatfix,twocolumn]{revtex4-1}
\usepackage{parskip} 
\usepackage{tikz} 
\usepackage{graphicx}
\usepackage{subfigure}
\usepackage{amsmath}
\usepackage{bbold}
\usepackage{slashed}
\usepackage{amssymb}
\usepackage{braket}
\usepackage{array}
\usepackage[colorlinks]{hyperref}
\usepackage{float}
\usepackage[margin=1in]{geometry}

\usepackage{natbib}

\begin{document}
\preprint{APS/123-QED}

\title { Three-dimensional $\mathbb{Z}$ topological insulators without reflection symmetry }
\author{Alexander C. Tyner$^{1,2}$}
\email{alexander.tyner@su.se}
\author{Vladimir Juri\v ci\' c$^{3,1}$}
\email{vladimir.juricic@usm.cl}
\affiliation{$^1$Nordita, KTH Royal Institute of Technology and Stockholm University, Hannes Alfv\'ens v\"ag 12, 106 91 Stockholm, Sweden}
\affiliation{$^2$ Department of Physics, University of Connecticut, Storrs, Connecticut 06269, USA}
\affiliation{$^3$ Departamento de F\'isica, Universidad T\'ecnica Federico Santa Mar\'ia, Casilla 110, Valpara\'iso, Chile}

\date{\today}

\begin{abstract} 
\textbf{Abstract}: In recent decades, the Altland-Zirnabuer (AZ)  table has proven incredibly powerful in delineating constraints for topological classification of a given band-insulator based on dimension and (nonspatial) symmetry class, and has also been  expanded by considering additional crystalline symmetries. Nevertheless, realizing a three-dimensional (3D), time-reversal symmetric (class AII) topological insulator (TI) in the absence of reflection symmetries, with a classification beyond the $\mathbb{Z}_{2}$ paradigm remains an open problem. In this work we present a general procedure for constructing such systems within the framework of  projected topological branes (PTBs). In particular, a 3D projected brane  from a ``parent"  four-dimensional topological insulator exhibits a $\mathbb{Z}$ topological classification,  corroborated through its response to the inserted bulk monopole loop.  More generally, PTBs have been demonstrated to be an effective route to performing dimensional reduction and embedding the topology of a $(d+1)$-dimensional ``parent" Hamiltonian in $d$ dimensions, yielding lower-dimensional topological phases beyond the AZ classification without additional symmetries. Our findings  should be relevant for the metamaterial platforms, such as photonic and phononic crystals, topolectric circuits, and designer systems.
\end{abstract}

\maketitle
\par 
\section{Introduction}
\par
Despite the rapid developments in our understanding and diagnosis of topological phases of matter in recent decades, the classification rules provided by the Altland-Zirnbauer (AZ) table have remained steadfast when only non-spatial  symmetries, such as time-reversal and particle-hole, are considered\cite{RyuRMP,RevModPhys.83.1057,ryu2010topological,RyuLudwigPRB,PhysRevB.88.075142,PhysRevB.88.125129}. In this respect, a few known exceptions to the AZ table exist~\cite{MooreHopf,PhysRevLett.126.216404,PhysRevResearch.3.033045,das2023hybrid,tyner2022dipolar}, perhaps most famous is the Hopf-insulator, which  achieves classification beyond the AZ table in three dimensions. 

\par This rather unexpected exception to the tenfold periodic table of topological insulating phases has naturally led to questions on what, if any, other exceptions may exist to prompt a reexamination of this paradigm. Moreover, rapid advances in meta-materials and engineered (designer) systems~\cite{zheludev2012metamaterials,wang2009observation,hafezi2013imaging,rechtsman2013photonic,nash2015topological,susstrunk2015observation,PhysRevX.5.021031,peterson2018quantized,PhysRevLett.122.233902,PhysRevLett.122.233903} for realizing the exotic physics of such exceptional systems in an experimental setting provides additional motivation to answer this question. 
\par
\begin{figure}[t!]
    \centering
\includegraphics[width=6cm]{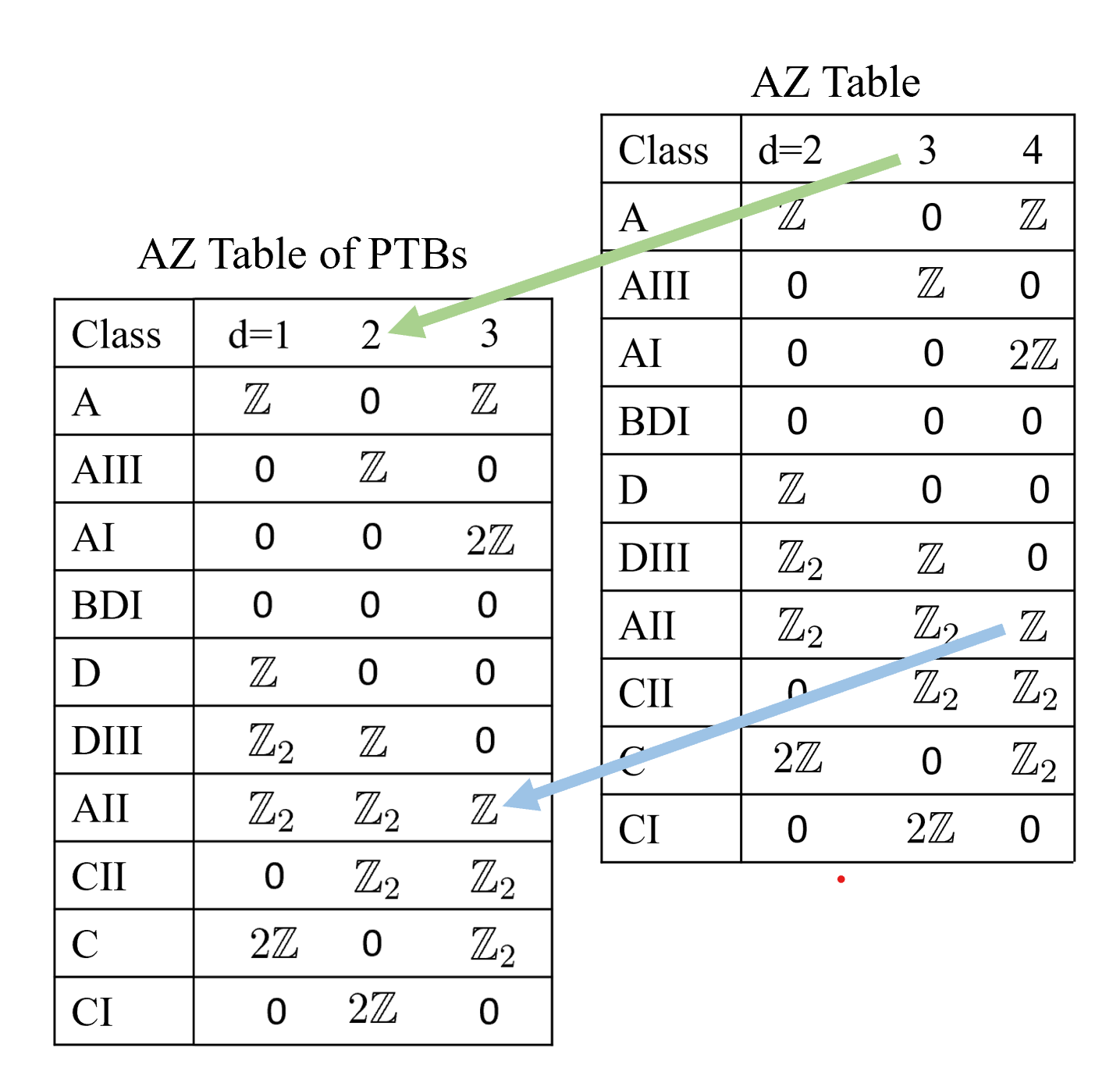}
  \caption{Periodic table of topological invariants for projected branes. Topological classification for each class in $d$-dimensions is inherited from the prescribed classification for $(d+1)$-dimensions in the Altland-Zirnbauer table. Therefore, PTBs can realize three-dimensional (3D) $\mathbb{Z}$ topological insulator protected purely by time-reversal symmetry. Additional example of the $\mathbb{Z}$-classified two-dimensional projected branes in class AIII with the chiral (unitary particle-hole) symmetry is discussed in the Supplementary Materials~\cite{SM}. This analysis shows that the 2D projected brane in this class inherits the $\mathbb{Z}$ topological invariant from the parent 3D state, consistent with the periodic table for projected branes displayed here.  }
    \label{fig:AZTablePTBs}
\end{figure}

 
{\color{black} In Ref.~\cite{panigrahi2022projected}, projected topological branes (PTBs) were introduced as a robust pathway to performing dimensional reduction of a lattice tight-binding model while preserving the bulk topology. This construction   was exemplified on  one- and two-dimensional PTBs obtained  from two-dimensional Chern insulators and three-dimensional (3D) Weyl semimetals, respectively.  
In this work, we generalize this method further, placing special emphasis on systems beyond the physical three spatial dimensions.} In particular, we demonstrate that topological classification of the $(d+1)$ dimensional system is preserved in the dimensional reduction procedure, realizing a $d-$dimensional topological brane. This procedure thereby offers a route to traverse the AZ table, as shown in Fig.~\ref{fig:AZTablePTBs}, and realize topological states in lower dimensions which would otherwise not be permitted. 

\begin{figure*}
\centering

\includegraphics[width=15cm]{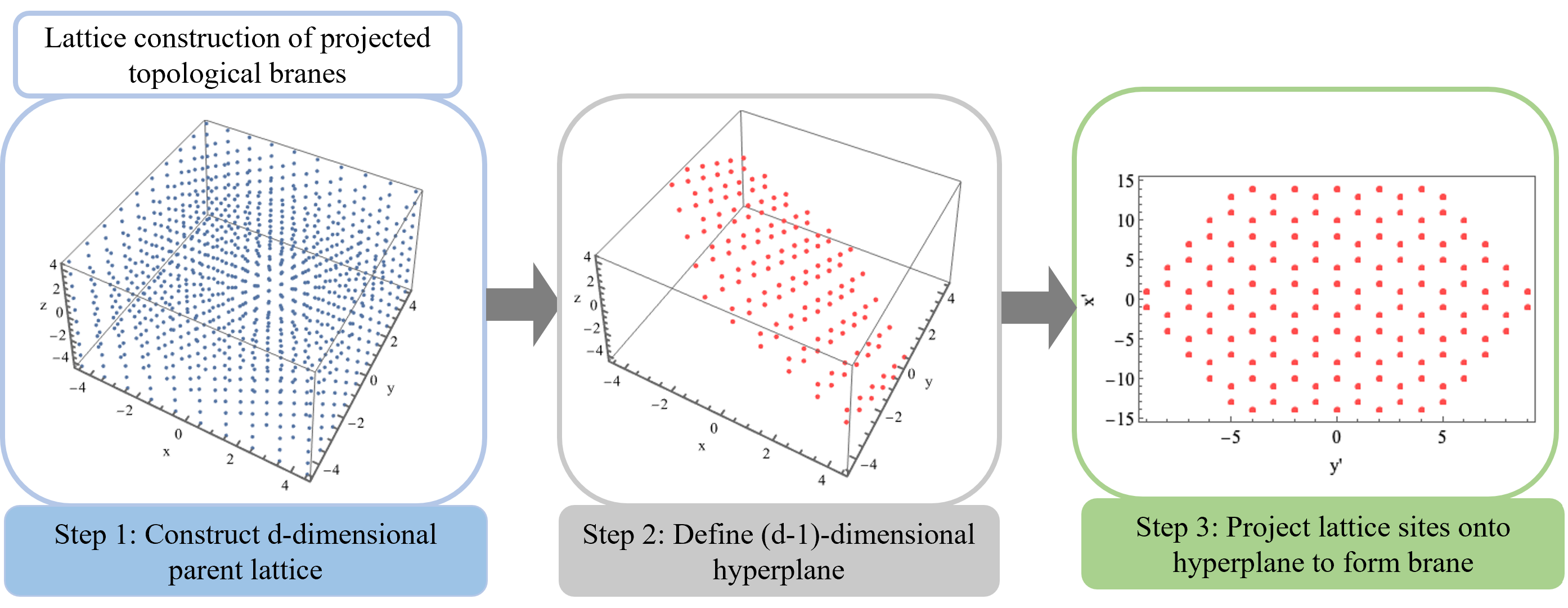}

\caption{Flow diagram detailing construction of $(d-1)$-dimensional projected brane from parent $d$-dimensional lattice. For clarity, we here show the example of parent  $10 \times 10 \times 10$ cubic lattice in three-dimensions with unit lattice constant along each direction.}
\label{fig:PTBCon}
\end{figure*}

\par 
For clarity, we focus on time-reversal symmetric (class AII) insulators in three dimensions as these represent a majority of real, spinful, condensed-matter systems. In the presence of reflection symmetry, they are characterized by an integer topological invariant, therefore  accommodating  $\mathbb{Z}$ classification~\cite{PhysRevB.88.075142,RyuRMP}. However, for general time-reversal insulators lacking additional symmetry, the AZ table limits topological classification to only a parity, $\mathbb{Z}_{2}$, invariant. By contrast, in four spatial dimensions class AII insulators  admit a $\mathbb{Z}$ invariant.
\par 
By forming 3D PTBs from four-dimensional topological insulators in class AII, we can thus provide a general principle for achieving the insulators with a  $\mathbb{Z}$ topological invariant  in three dimensions, thereby going beyond the AZ table without introducing  symmetry constraints. In order to prove that we have preserved the topological nature of the parent system, we utilize known real-space probes of bulk topology, namely electromagnetic vortices, monopoles, and monopole loops for parent Hamiltonians in $d=2,3,4$, respectively\cite{QiSpinCharge,SpinChargeVishwanath,slager2012,MESAROS2013977,Wang_2010,tyner2020topology,tynerbismuthene,rosenberg2010witten}. In particular, the spectrum of the prototypical, four-dimensional (4D) time-reversal symmetric model with a monopole-loop inserted in the fourth dimension yields a number of induced mid-gap modes, $N_0=2|\mathcal{C}_2|$, that is in correspondence with the second Chern number, $\mathcal{C}_2$, characterizing the 4D topological state. Remarkably, this number of mid-gap modes remains invariant when the 3D PTB is constructed out of the 4D parent state, therefore demonstrating the  realization of  $\mathbb{Z}$ classified time-reversal invariant topological insulators in $d=3$.

\par 
\section{Projected topological branes}
\par
{\color{black}To construct the PTB, we employ the method based on the Schur complement~\cite{panigrahi2022projected}. Consider a lattice tight-binding model defined in $d$-dimensions for a system of $N^d$ lattice sites, which we refer to as the parent system. The real-space Hamiltonian for the parent system can be written in the block form as 
\begin{equation}
    H=\begin{bmatrix}
    H_{11} & H_{12}\\
    H_{21} & H_{22}
    \end{bmatrix}.
\end{equation} 
The Hamiltonian for the PTB then takes the form~\cite{panigrahi2022projected},
\begin{equation}\label{eq:PTB}
    H_{PTB}=H_{11}-H_{12}H_{22}^{-1}H_{21},
\end{equation} 
where $H_{11}$ corresponds to the real-space Hamiltonian of lattice sites on the $(d-1)$-dimensional brane. By contrast, $H_{22}$ is the real-space Hamiltonian for  the remaining subsystem formed by  lattice sites which \emph{do not} constitute the $(d-1)$-dimensional brane.  It then follows that the off-diagonal piece in the Hamiltonian~\eqref{eq:PTB}, $H_{12}=H_{21}^\dagger$, is the coupling between the  PTB and the remaining subsystem in the parent $d$-dimensional lattice tight-binding model.}
\par
While the form of Eq. \eqref{eq:PTB} is insensitive to choice of dimension, we detail the generalization of this algorithm for constructing $(d-1)$-dimensional branes from $d$-dimensional parent cubic lattices. Generalization of this procedure for $d$-dimensional parent systems relies on specifying a $(d-1)$-dimensional hyperplane, the equation of which takes the form, 
\begin{equation}\label{eq:hyper}
    \sum_{j=1}^{d} \alpha_{j}x_{j}=\beta,
\end{equation}
where $\alpha_{j}$ and $\beta$ are real numbers. A lattice site, $i$, projected onto the $(d-1)$-dimensional PTB, obeys the relation,
\begin{equation}\label{eq:const}
   \frac{ |\sum_{j=1}^{d}\alpha_{j} x_{j,i}+\beta|}{\sqrt{\sum_{j=1}^{d}\alpha_{j}^2}}<\frac{1}{\sqrt{2}a},
\end{equation}
where $a$ is the lattice constant. As an illustrative example, we choose a parent cubic system in $d=3$ dimensions of size $10 \times 10 \times 10$. We then select the parameters for our plane, $\alpha_{j=1,2,3}=1$ and $\beta=1/100$. The lattice sites which make up the PTB are colored in red in the middle panel of Fig.~\ref{fig:PTBCon}. In order to visualize the PTB, we map each lattice site in red onto the projected plane at the point nearest to the lattice site. The result is shown in the right panel of Fig.~\ref{fig:PTBCon}, demonstrating that the PTB is a hexagonal system as expected given the orientation of the plane, perpendicular to the $(111)$ axis of the cube. 
\par
{\color{black}  At this point we remark that an infinite number of choices for the parameters which define the hyperplane in Eq. \eqref{eq:hyper} may be made. Tuning such parameters may allow to tune the band topology of the corresponding PTB. For example, a two-dimensional PTB 
may not inherit the bulk topology of the parent model if the selected hyperplane extends only in a two-dimensional subspace corresponding to a single layer  of the parent 3D  lattice with stacked two-dimensional layers. In this work we continuously use a hyperplane extending along the body diagonal such that a genuine dimensional reduction is performed and ensures that the PTB inherits the bulk band topology of the parent system, thus avoiding previously mentioned trivial projection.} Having established a definite procedure for construction of projected branes in $d$-dimensions, we consider a 4D, parent tight-binding model for demonstration of topology beyond the AZ table. 
\par 
{\color{black}\section{Construction and analysis of three-dimensional PTB}}
{\color{black}\subsection{Bulk topology of four-dimensional parent model}}
\par 
We explicitly consider a 4D generalization of the Bernevig-Hughes-Zhang topological insulator\cite{bernevig2006quantum,Qi2008} on a cubic lattice. The Bloch Hamiltonian takes the form, $H(\mathbf{k})= \sum_{j=1}^{5}d_{j}(\mathbf{k})\Gamma_{j}$. Employing the basis, 
\begin{equation}\label{eq:normalbasis}
    \Gamma_{j=1,2,3}=\tau_{1}\otimes \sigma_{j},\; \Gamma_{4}= \tau_{2}\otimes \sigma_{0},\; \Gamma_{5}= \tau_{3} \otimes \sigma_{0},
\end{equation}
where $\tau_{0,1,2,3}(\sigma_{0,1,2,3})$ are the $2 \times 2$ identity matrix and three Pauli matrices respectively, acting on the orbital (spin) degrees of freedom, the vector $\mathbf{d}(\mathbf{k})$  reads, 
\begin{multline}\label{eq:4D_TB}
d_{j=1,2,3,4}(\mathbf{k})=t_{p}\sin k_{j},\; \\ d_{5}(\mathbf{k})= t_{s}(\Delta- \eta_{1}\sum_{j=3}^{4}\cos k_{j}-\eta_{2}\sum_{j=1}^{2}\cos k_{j}\\-\eta_{3}\cos k_{1}\cos k_{2})\Gamma_{5}.
\end{multline}
We have set the lattice constant to unity for simplicity, $t_{p,s}$ have units of energy and $\Delta,\; \eta_{1,2,3}$ are dimensionless, real, non-thermal band parameters used for driving topological phase transitions. 
\par 
Time-reversal symmetry $\mathcal{T}$, is generated by $\mathcal{T}^{\dagger}H^{*}(-\mathbf{k})\mathcal{T}=H(\mathbf{k})$, where $\mathcal{T}=i\tau_{0}\otimes \sigma_{2}$, such that $\mathcal{T}^{2}=-1$, placing the Hamiltonian in class AII. As such, it supports $\mathbb{Z}$ topological classification via calculation of the second Chern number, $\mathcal{C}_{2}$\cite{zhang2001four,Qi2008}. The second Chern number is efficiently computed for the model at hand as, 
\begin{equation}\label{eq:winding}
\mathcal{C}_2= \frac{3}{8\pi^{2}} \;  \int d^{4}k\; \epsilon^{abcde} \hat{d}_{a} \frac{\partial \hat{d}_{b}}{\partial k_x} \frac{\partial \hat{d}_{c}}{\partial k_y}\frac{\partial \hat{d}_{d}}{\partial k_z}\frac{\partial \hat{d}_{e}}{\partial k_w}.
\end{equation}
We will consider three main parameter sets, (A) $\Delta =3.5, \; (\eta_{1},\eta_{2},\eta_{3})=(1,1,0)$, (B) $\Delta =0.5, \; (\eta_{1},\eta_{2},\eta_{3})=(1,1,0)$, and (C) $\Delta =0.5, \; (\eta_{1},\eta_{2},\eta_{3})=(1,0,1)$. These phases support $\mathcal{C}_{2}=-1, +3, +4$ respectively. 

\begin{figure}
    \centering
    \includegraphics[width=8cm]{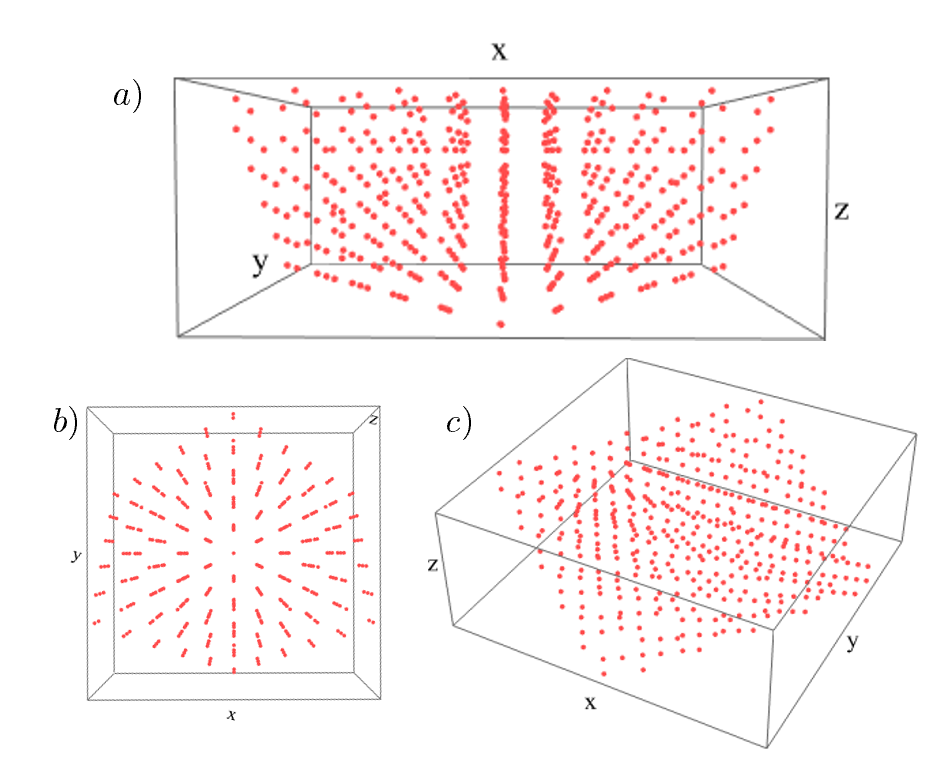}
    \caption{Three-dimensional projected brane formed from the four-dimensional hypercubic crystal. (a) Side, (b) top, and (c) corner perspective on projection of lattice sites in four-dimensional parent lattice onto three-dimensional hyperplane, forming the brane.}
    \label{fig:4DPTBLattice}
\end{figure}

\begin{figure*}
\centering
\subfigure[]{
\includegraphics[scale=0.58]{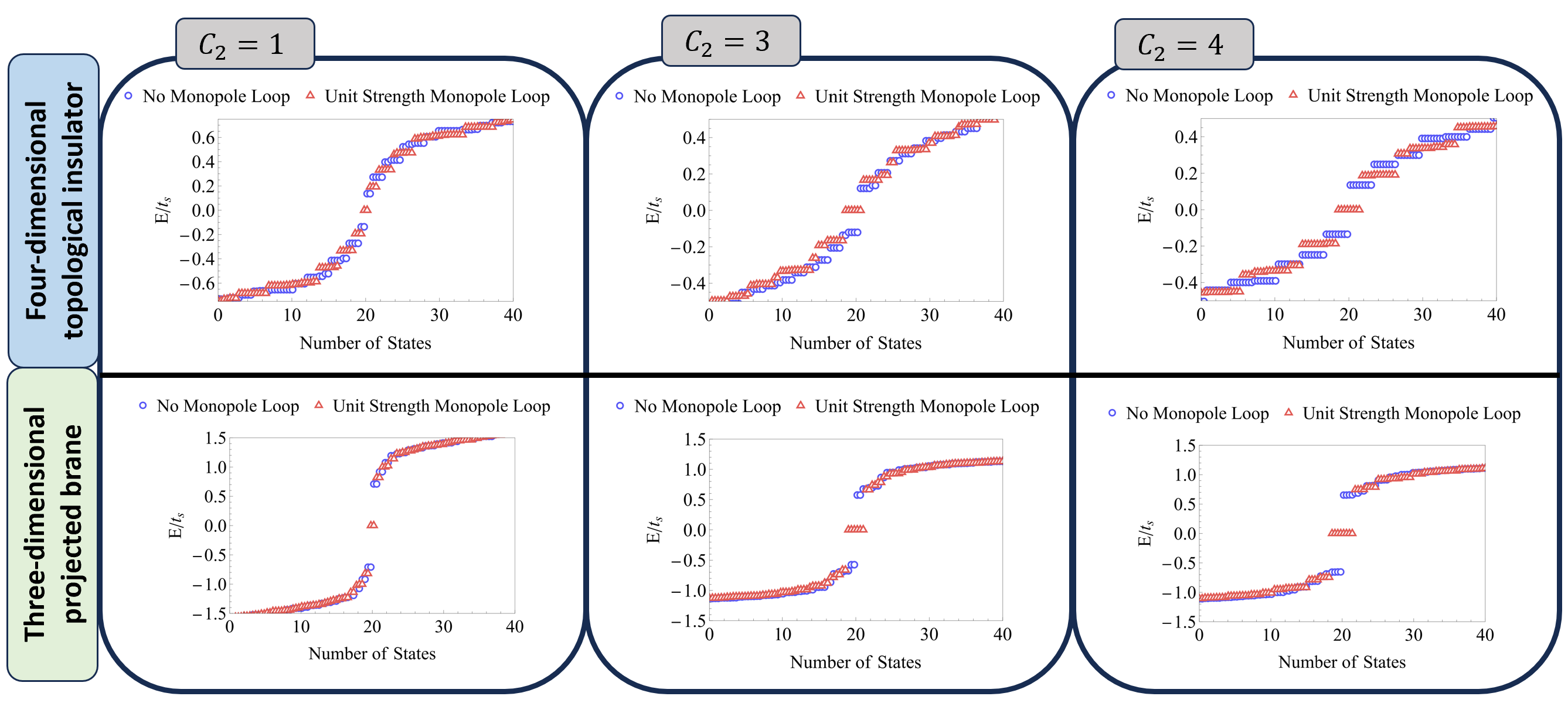}
\label{fig:4DTI1}}
\caption{Monopole loop as a bulk probe of the three-dimensional $\mathbb{Z}$ topological insulator realized on a projected brane from its  four-dimensional ``parent'' topological state. Above:
Close-to-zero-energy states for the tight-binding model given by Eq.~\eqref{eq:4D_TB}  for a $10 \times 10 \times 10 \times 10$ hypercubic lattice. Open boundary conditions are imposed along the $x,y,z$ directions with periodic boundary conditions along the (fourth) $w$ direction. The spectra are shown with and without a unit strength monopole loop along the $w$ direction. The number of zero-energy states is in direct correspondence with the second Chern number for phase A, B, and C, ($C_{2}=1,3,4$). Below: Spectra of three-dimensional projected topological brane formed from the four-dimensional topological insulator. The number of zero modes remains invariant, demonstrating that the brane has inherited the topological character of the four-dimensional phase.}
\label{fig:4DPTB}
\end{figure*}
\par 
{\color{black}\subsection{Topological analysis of three-dimensional PTB}}
\par 
Utilizing the parent, 4D Bloch Hamiltonian detailed in Eq.~\eqref{eq:4D_TB}, we will now construct the 3D PTB following the previously detailed procedure, fixing $\alpha_{j=1,2,3,4}=1$ and $\beta=0.01$. The lattice sites projected onto the 3D hyperplane are shown in Fig.~\ref{fig:4DPTBLattice}. In order to diagnose the bulk topology of the PTB we will utilize a real-space probes in terms of defects. In particular, singular magnetic probes have been proven to be ideal in detecting topology in real-space through the emergence  of bound states. Famously, in two-dimensions insertion of magnetic vortices has been used to determine spin-Chern number ($\mathcal{C}_{s}$), with the number of mid-gap vortex-bound modes ($N_{VBM}$), following the relationship $N_{VBM}=2\mathcal{C}_{s}$\cite{QiSpinCharge,SpinChargeVishwanath,slager2012,MESAROS2013977,tynerbismuthene}. Furthermore, in three dimensions, magnetic monopoles through  the number of monopole bound mid-gap modes  was employed to diagnose the bulk-invariant\cite{witten1979dyons,yamagishi1983fermion,yamagishi1983fermion2,shnir2006magnetic,rosenberg2010witten,zhao2012magnetic}. In 4D systems, a natural extension is the monopole loop, whereby a unit strength magnetic monopole is placed in each plane parameterized by the conserved momenta in the fourth dimension, $k_{4}$. Notice that the high-symmetry values of $k_{4}$, namely $k_{4}=0,\pm \pi$, represent distinct 3D topological insulators. Importantly, the emergent 3D topological insulator defined at these planes supports a chiral (unitary particle-hole) symmetry defined as, $S^{-1}HS=-H$ where $S=\Gamma_{4}$. This emergent chiral symmetry at the high-symmetry locations $k_{4}=0,\pm \pi$, in turn, pins to zero energy the surface and monopole bound states induced by the monopole-loop. Finally, even through the chiral symmetry can be broken, as  allowed for time-reversal symmetric insulators (class AII), the number of bound states remains invariant, consistent with the $\mathbb{Z}$ classification, but they are simply shifted to finite energy values.

\par 
We insert the monopole loop into the bulk under open-boundary conditions along the $x,y,z$ directions and periodic boundary conditions along the $w$ direction, by employing the singular, north-pole gauge
 \begin{eqnarray}
 \mathbf{A}(\mathbf{r}_i)= \frac{g}{r_i} \cot \frac{\theta_i}{2} \; \hat{\phi}_i=g \; \frac{-y_i \hat{x} + x_i \hat{y}}{r_i(r_i+z_i)}, 
 \end{eqnarray}
where $i$ index lattice sites and we fix $g=1$ to specifically consider the case of a unit-strength monopole loop.
\par 
The results of inserting the monopole loop into the parent Hamiltonian for phase (A), (B), and (C) are shown in the top panel of Fig.~\ref{fig:4DPTB}, detailing that the number of mid-gap zero modes in each phase precisely follows the relationship, $N_{0}= 2|\mathcal{C}_{2}|$. Having established this relationship for the parent system, we perform the projection to construct the topological brane. We carry out this process both with and without  the monopole loop inserted, maintaining identical boundary conditions utilized for examining the four-dimensional system. 
\par 
Solving for the spectra of the projected brane in each phase, we find the results shown in the bottom panel of Fig.~\ref{fig:4DPTB}. Remarkably, the number of mid-gap zero modes in each phase under insertion of the monopole loop remains invariant, thereby demonstrating that the topology of the parent 4D topological state is inherited by the 3D PTB.

\par 
\section{Summary and outlook}
\par
In this work we have demonstrated that the PTBs offer a route to perform dimensional reduction of a parent Hamiltonian in $(d+1)$ dimensions to a $d-$dimensional PTB, while preserving the bulk topological invariant. Importantly, this allows for the construction of lattice tight-binding models for which the bulk topology goes beyond the ten-fold classification scheme based on the AZ table. PTBs thus fall under the category of symmetry non-indicative phases, of which other known examples include the Hopf insulator. As the projected branes can be constructed through lattice tight-binding models, opportunities exist to realize these systems in engineered metamaterial systems, including photonic, phononic and topolectric systems. Furthermore, the designer quantum materials offer another route to experimentally test our proposal. {\color{black} While routes to constructing synthetic dimensions exist\cite{4DColdAtoms,4DHall,zilberberg2018photonic}, PTBs offer an alternate route to exploring such physics without the additional requirement of synthetic dimensions.} This important direction for physical realization of the exotic properties will be pursued in a subsequent work. We also expect that new studies accounting for the effects of disorder will further corroborate the robustness of the projected branes  arising from their topological nature. Finally, higher-dimensional crystalline dislocations, being related to translations in extra dimensions, should provide a refined classification of the 3D  $\mathbb{Z}$ projected branes, which we plan to study in future. 

\par
\textbf{Acknowledgments:} {We are thankful to Bitan Roy for useful discussions and critical reading of the manuscript.  V. J. acknowledges the support of  the Swedish Research Council Grant No. VR 2019-04735  and   Fondecyt (Chile) Grant No. 1230933. Nordita is supported in part by NordForsk.}

\par
\textbf{Contributions} A.C.T. and V.J. conceived the project. A. C. T. carried out the calculations. V.J. and A. C. T. wrote the paper. \\

\textbf{Competing Interests} The authors declare that they have no competing financial interests. \\

\textbf{Data Availability} The datasets used and/or analysed during the current study available from the corresponding author on reasonable request.

\textbf{Corresponding authors} Correspondence and requests for materials should be addressed to A.C.T. (email: alexander.tyner@su.se) or V.J. (email: vladimir.juricic@usm.cl). 


\bibliographystyle{apsrev4-1}
\nocite{apsrev41Control}
\bibliography{ref.bib}

\end{document}